\documentclass[preprintnumbers,showpacs,amsmath,amssymb,floatfix,prd,twocolumn,superscriptaddress,nofootinbib]{revtex4}
\usepackage{graphicx}
\usepackage{epsfig}
\usepackage{bm}
\usepackage{amsfonts}

\begin{document}

\title{Preventing eternality in phantom inflation}

\author{Chao-Jun Feng}
\email{fengcj@shnu.edu.cn} \affiliation{Shanghai United Center for
Astrophysics (SUCA), \\ Shanghai Normal University,
    100 Guilin Road, Shanghai 200234, People's Republic of China}

\author{Xin-Zhou Li}
\email{kychz@shnu.edu.cn} \affiliation{Shanghai United Center for
Astrophysics (SUCA),  \\ Shanghai Normal University,
    100 Guilin Road, Shanghai 200234, People's Republic of China}

\author{Emmanuel N. Saridakis}
 \email{msaridak@phys.uoa.gr}
 \affiliation{College of Mathematics
and Physics,\\ Chongqing University of Posts and
Telecommunications, Chongqing, 400065, People's Republic of China }

\begin{abstract}

We have investigated the necessary conditions that prevent phantom inflation from being eternal. Allowing additionally
for a nonminimal coupling between the phantom field and gravity, we present the slow-climb requirements, perform an
analysis of the fluctuations, and finally we extract the overall conditions that are necessary in order to prevent
eternality. Furthermore, we verify our results by solving explicitly the cosmological equations in a simple example of
an exponential potential, formulating the classical motion plus the stochastic effect of the fluctuations through
Langevin equations. Our analysis shows that phantom inflation can be finite without the need of additional exotic
mechanisms.
 \end{abstract}


 \maketitle

\section{Introduction}

After almost three decades of extensive research, inflation is now
considered to be a crucial part of the cosmological history of the
Universe \cite{inflation}, having affected indelibly its
observational features. Introducing a scalar field, the inflaton,
and a suitable potential, one can make various scenarios of
inflation realization in conventional, as well as in higher-dimensional frameworks \cite{Lindebook,LidseyRMP,Baumann:2009ni}.
Additionally, one could generalize the aforementioned paradigm,
allowing for a nonminimal interaction of the scalar field with
gravity \cite{nonminimal0}, since nonminimal inflation could
improve the obtained perturbation spectrum
   \cite{miaoLi,recent nonminimal}.

One important subject that has to be  addressed in this paradigm is that of the exit from the inflationary epoch, that
is to examine whether inflation can be eternal or not. In particular, in the new inflation scenario it was shown that
the procedure could be eternal since the ``false'' vacuum (in which the field lies during inflation) is never dominated
by the ``true'' one (the approach of which causes the end of inflation) \cite{steinhardt-nuffield,Vilenkin:1983xq}.
Additionally, even in advanced scenarios, such as the chaotic inflation, where there is no false vacuum state,
slow-roll eternality is also possible \cite{Linde:1986fd} due to a different mechanism. In particular, in this
model-subclass the inflaton is classically rolling down its potential slope, however the quantum fluctuations can
conditionally drive it upwards and thus inflation will never end \cite{recent pro eternal,Linde:2007fr,Guth:2007ng}.
Thus, one must in general examine the conditions for the realization of eternality \cite{Guth:2000ka}.

An interesting class of inflation scenarios
\cite{Piao:2004tq,Lidsey:2004xd,Nojiri:2005pu,Capozziello:2005tf,GonzalezDiaz:2004df,Wu:2006wu,Piao:2007ne,Elizalde:2008yf}
is achieved through the use of phantom fields \cite{phant0}, inspired by the wide use of such fields to explain the
late-time universe acceleration \cite{phant1}. The simplest realization of phantom fields is the use of a negative
kinetic term in the Lagrangian, but this could lead their quantum theory to be problematic, due to the causality and
stability problems and the possible spontaneous breakdown of the vacuum into phantoms and conventional particles
\cite{CHT, Cline:2003gs}. However, one could consider that the phantom fields arise through an effective description of
a nonphantom fundamental (probably higher-dimensional) underlying theory, consistently with the basic requirements of
quantum field theory \cite{quantumphantom0}. Indeed actions with phantomlike behavior may arise in supergravity
\cite{N}, scalar tensor gravity \cite{BEPS}, higher derivative gravity \cite{P}, braneworld \cite{SS}, k-field
\cite{ADM}, stringy \cite{Frampton:2002tu} and others scenarios \cite{Chimento:2003qy, Stefancic:2003bj}.

The peculiar nature of phantom fields requires the inflation
paradigm to be suitably redesigned. In particular, since phantoms
behave inversely in potential slopes, climbing up along them, in
order to avoid an early-time Big Rip singularity \cite{BigRip}, one
must use potentials with maxima instead of minima, and the slow-roll parameters are replaced by the ``slow-climb'' ones
\cite{Sami:2003xv}. However, even with potentials bounded from
above, the problem of eternal inflation still exists, and one
should examine in detail the possible exits from the inflationary
epoch
\cite{Piao:2004tq,GonzalezDiaz:2004df,Wu:2006wu,Piao:2007ne,Elizalde:2008yf}.

In the present work we are interested in investigating the
necessary conditions that prevent phantom inflation from being
eternal, going beyond the basic requirements of bounded-from-above
potentials. In particular, we examine whether the quantum
fluctuations could affect the classical motion towards the
potential maximum, preventing inflation to the end. Furthermore, in
order to be general we allow for a nonminimal coupling of the
phantom field with gravity, since this interaction could also
affect the eternality conditions, similarly to the canonical case
\cite{Feng:2009kb}.

The plan of this work is the following: In Sec. 
\ref{phant.slowroll} we present the phantom-inflation scenario
under the conditions of slow-climb. In Sec. \ref{flu inf} we
perform a fluctuation analysis and we extract the conditions for
preventing eternality, while in Sec. \ref{Langanal} we verify
our results by solving explicitly the Langevin equations for the
cosmological evolution in a simple example. Finally, Sec.
\ref{conclusions} is devoted to the summary of the obtained
results.

\section{Phantom slow-climb inflation}
\label{phant.slowroll}

Let us present briefly the cosmological scenario of nonminimal
phantom inflation \cite{Elizalde:2008yf}, focusing on the
conditions required for its long-time, efficient duration. The
action of a universe constituted by a phantom scalar field
$\varphi$ is given by
\begin{equation}\label{action}
    S = \int d^4x \sqrt{-g}\left[\frac{R}{2}+  \frac{1}{2}g^{\mu\nu}\partial_\mu\varphi\partial_\nu\varphi -
    V(\varphi)  - \frac{1}{2}f(R)\varphi^2\right],
\end{equation}
with $V(\varphi)$ the corresponding potential, and where for
simplicity we have set $8\pi G = M_p^{-2} = 1$. As usual $R$ is
the Ricci scalar, and $f(R)$ is the function describing the
coupling of the phantom field to gravity. Throughout this work we
consider a flat Friedmann-Robertson-Walker geometry with the
unperturbed metric
\begin{equation}\label{metric}
    ds^2 = -dt^2 + a^2(t)\left(dx^2 + dy^2 + dz^2\right),
\end{equation}
with $a(t)$ the scale factor and $t$ the comoving time. Thus,
defining the Hubble parameter as $H \equiv \dot a/a$ the
 scalar curvature reads:
\begin{equation}\label{scalar cur}
    R = 6\left(\dot H + 2H^2\right),
\end{equation}
where the dot denotes the derivative with respect to $t$.

Variation of the action (\ref{action}) leads to the two Friedmann
equations
\begin{equation}
  3H^2 = -\frac{1}{2}\dot\varphi^2 + V + \frac{1}{2}f\varphi^2 + 3H^2\left[ \frac{d}{dt}\left(\frac{f'\varphi^2}{H}\right) -
    f'\varphi^2\right]  \label{Fried1}
    \end{equation}
 \begin{equation}
 -2\dot H = -\dot\varphi^2 + H^3 \frac{d}{dt}\left(\frac{f'\varphi^2}{H^2}\right)- \frac{d^2}{dt^2}\bigg(f'\varphi^2\bigg) ,
    \end{equation}
where the prime denotes differentiation with respect to the
corresponding argument, that is $f' = df/dR$ and $V' =
dV/d\varphi$. Additionally, the evolution equation for the scalar
field writes:
\begin{equation}\label{eomp1}
    \ddot \varphi + 3H \dot\varphi - V' - f\varphi = 0.
\end{equation}

As we discussed in the Introduction, the phantom fields require
potentials bounded from above, since they climb upwards the
potential slopes. Therefore, in order to acquire a long-time
inflation in phantom cosmology we impose the following slow-climb
conditions \cite{Sami:2003xv}
\begin{equation}
   |\dot H| \ll H^2 \,, \quad  |\ddot \varphi| \ll 3H|\dot\varphi| ,
\end{equation}
which corresponds to the slow-roll conditions of canonical
inflation \cite{Lindebook}. After some algebra, and assuming
potential domination ($|-\frac{\dot\varphi^2}{2}|\ll V$) to
simplify the calculations, the slow-climb conditions write as
\begin{equation}\label{cond1}
 |f'\varphi^2 \dot H| \ll V
\,,\quad \bigg|V'' + f - \frac{3Hf'\dot
R}{f}\frac{f\varphi}{V'+f\varphi}\bigg| \ll 9H^2 .
\end{equation}
 Furthermore, in the usual case where $f$ is a monomial of
$R$, for instance $f\sim R^n$, we obtain $3H (\log f)'\dot R
\approx 6n\dot H $. Therefore, if $|f\varphi|\ll |V'|$ or
$|f\varphi|\gg |V'|$, the third term on the left hand side of the
second equation in (\ref{cond1}) can be neglected. Thus, the
aforementioned expressions are simplified to
\begin{equation}\label{cond2}
|f'\varphi^2 \dot H| \ll V \,, \quad \bigg|V'' + f \bigg| \ll 9H^2\,.
\end{equation}
In summary, under these slow-climb conditions, the first Friedmann
equation  (\ref{Fried1}) becomes
\begin{equation}
3H^2 = \frac{1}{2}\left(f-6f'H^2\right)\varphi^2 + V \,, \label{Fried1b}\\
\end{equation}
while the phantom field equation of motion, for the two examined
limiting cases, is simplified as
\begin{eqnarray}
  3H \dot\varphi &=&  V' \,,  \quad |f\varphi|\ll |V'|\quad(\text{Case I}~ ) \label{eomp21}\,,\\
  3H \dot\varphi &=&  f\varphi \,, \quad |f\varphi|\gg |V'|\quad(\text{Case II})\label{eomp22}\,.
\end{eqnarray}

At this stage we introduce the standard
 dimensionless slow-climb parameters as \cite{Piao:2004tq,Wu:2006wu}
\begin{equation}\label{slow para}
   \epsilon = -\frac{\dot H}{H^2} \,, \quad \eta = M_p^2\frac{V''}{V} \,,
\end{equation}
and following \cite{Feng:2009kb} we define a new dimensionless
slow-climb parameter
\begin{equation}
 \label{slow para2}
    \Delta \equiv M_p^2 \frac{f}{V}
\end{equation}
to account for the nonminimal coupling, where we have recovered
the Planck mass to indicate that these parameters are indeed
dimensionless. Using these parameters, the slow-climb conditions
(\ref{cond2}) become
\begin{equation}\label{cond3}
    \epsilon \Delta \varphi^2\ll1 \,, \quad \eta + \Delta \ll1,
\end{equation}
having also used for simplicity $f'\sim f/R$ although this is not
necessary. Therefore, if $\epsilon , \eta ,\Delta\ll1$, the
slow-climb conditions (\ref{cond2}) are indeed satisfied.

 In order to continue, we consider explicitly the usual ansatz for
$f(R)$ of the literature \cite{nonminimal0,recent nonminimal},
namely  $f = \xi R$, with $\xi$ the coupling parameter. Thus, the
Friedmann equation (\ref{Fried1b}) becomes
\begin{equation}\label{Fried mod1}
    3H^2 = \frac{V}{1-\xi\varphi^2} \,,
\end{equation}
and the slow-climb parameter $\Delta$ reads
\begin{equation}\label{del1}
    \Delta = \frac{2\xi(2-\epsilon)}{(1-\xi\varphi^2)}.
\end{equation}
As we see, $\Delta\ll1$ requires $\xi\ll1$ in the model at hand, a
condition which is usually satisfied in all nonminimal scenarios.
Finally, differentiating the Friedmann equation (\ref{Fried1b}),
we deduce that in the case $|f\varphi|\ll |V'|$ (Case I), that is
$ \Delta \varphi^2 \ll1$, the slow-climb parameter $\epsilon$
becomes
\begin{equation}
    \epsilon  = -\frac{V'^2}{2V^2} \bigg[1-\left(1-\frac{2V}{V'\varphi}\right)\xi\varphi^2\bigg] \,,
\end{equation}
while in the case of $|f\varphi|\gg |V'|$ (Case II), i.e. $ \Delta
\varphi^2 \gg1$, it becomes
\begin{equation}
    \epsilon  =- \frac{f\varphi V'}{2V^2} \bigg[1-\left(1-\frac{2V}{V'\varphi}\right)\xi\varphi^2\bigg] \,.
\end{equation}
 Note that in the latter case the condition $ \Delta \varphi^2 \gg1$ requires the field values to be large
and therefore, without loss of generality, in the following we
consider large-field inflation.

\section{Fluctuations and conditions for preventing eternality}
\label{flu inf}

In the previous section we extracted the basic conditions for an efficient long time, but noneternal phantom
inflation. However, as we discussed in the Introduction, even if one manages to stop inflation at the classical level
using suitable potentials, the backreaction of the metric plus the inflaton's quantum fluctuations on the background
space-time could make the inflaton field follow a Brownian motion in which half of the time the inflaton field in a
given domain will jump downwards, instead of drifting up to the potential. Thus, the necessary conditions for preventing
eternality in phantom inflation will arise through examination of the overall effects of the classical behavior plus
the fluctuations.

In order to calculate the quantum fluctuation of the inflaton, we
expand the action (\ref{action}) to second order, since the action
approach guarantees the correct normalization for the quantization
of fluctuations. It is convenient to work in the Arnowitt-Deser-Misner (ADM) formalism and
write the metric as
\begin{equation}
    ds^2 = -N^2dt^2 + h_{ij}(dx^i + N^idt)(dx^j + N^jdt) \,,
\end{equation}
where $N$ is the lapse function and $N^i$ is the shift vector.
Note that such perturbations have been studied in a different
framework, for the minimal case, in \cite{Piao:2004tq,Wu:2006wu}.

The action (\ref{action}) becomes
\begin{eqnarray}\label{action2}
    S = \frac{1}{2} \int dt dx^3 \sqrt{h}\left[NR^{(3)} +
    N^{-1}\left(E_{ij}E^{ij}-E^2\right)\ \ \ \ \ \ \ \ \ \ \ \right.\nonumber\\
    \left.-N^{-1}\left(\dot\varphi - N^i\partial_i\varphi\right)^2 +
    Nh^{ij}\partial_i\varphi\partial_j\varphi-N\left(2V+f\varphi^2\right)\right],\
    \
\end{eqnarray}
where $h=\det h_{ij}$ and the symmetric tensor $E_{ij}$ is defined
as
\begin{equation}
    E_{ij} = \frac{1}{2}\bigg(\dot h_{ij}-\nabla_iN_j-\nabla_jN_i\bigg) \,, \quad E = E^i_{~i} \,.
\end{equation}
In  (\ref{action2}) $R^{(3)}$ is the three-dimensional Ricci
curvature, which is computed from the metric $h_{ij}$, and
$K_{ij}=E_{ij}/N$ is the extrinsic curvature. In the following we
work in the spatially-flat gauge and we neglect the tensor
perturbations. Thus, we write
\begin{equation}\label{flat gauge}
    \varphi(t,x) = \bar\varphi(t) + \delta\varphi(t,x) \,, \quad h_{ij}=a^2\delta_{ij} \,,
\end{equation}
where $\bar\varphi(t)$ is the background value of the scalar field
and $\delta\varphi$ is a small fluctuation around the background
value.

In the ADM formalism one can consider $N$ and $N^i$ as Lagrange
multipliers, and in order to obtain the action for $\xi$ one needs
to solve the constraint equations for $N$ and $N^i$ and substitute
the result back in the action. The equations of motion for $N^i$
and $N$ are the momentum and Hamiltonian constraints
\begin{eqnarray}
  \nabla_i\bigg[(1-f'\varphi^2)N^{-1}\left(E^i_j -
  \delta^i_jE\right)\bigg]\ \ \ \ \ \ \ \ \ \ \ \ \ \nonumber\\
   \ \ \ \ \ \ \ +N^{-1}\left(\dot\varphi-N^i\partial_i\varphi\right)\partial_j\varphi  = 0 \label{f constr1}
  \end{eqnarray}
  and
\begin{eqnarray}
  R^{(3)} -(1-2f'\varphi^2) N^{-2}\left(E_{ij}E^{ij}-E^2\right)\ \ \ \ \ \ \
  \
  \ \ \ \ \ \ \ \ \ \ \ \
 \nonumber\\
  +N^{-2}\left(\dot\varphi-N^i\partial_i\varphi\right)^2
  - 2V-f\varphi^2 +h^{ij}\partial_i\varphi\partial_j\varphi= 0
. \ \ \ \label{f constr2}
\end{eqnarray}
We now decompose $N^i$ into
\begin{eqnarray}
N^i = \partial^i\psi + N^i_T
\end{eqnarray}
 with
$\partial_iN^i_T=0$, and we define
\begin{eqnarray}
N_1\equiv  N-1,
 \end{eqnarray}
  where $N_1,N^{i}_T,\psi\sim
\mathcal{O}(\delta\varphi)$. Thus, inserting these expansions into
(\ref{f constr1}) and (\ref{f constr2}), we can obtain the
solutions up the first order in $\xi$. In particular, in the usual
case $f = \xi R$, we can derive the first order solutions
similarly to the Appendix of \cite{Feng:2009kb}. Simplifying the
notation using $\varphi$ to denote the background value
$\bar\varphi$, we finally acquire:
\begin{equation}\label{sol 11}
    N_1 = -\frac{\delta\varphi}{1-\xi\varphi^2} \left(\frac{\dot{\varphi}}{2H} + 2\xi\varphi \right)\,, \quad N^{i}_T =0 \,,
\end{equation}
and
\begin{eqnarray}\label{sol 12}
   (1-\xi\varphi^2) \partial^2\psi = -N_1\frac{\dot{\varphi}^2}{2H}+ \frac{\dot{\varphi}}{2H}\delta\dot\varphi
    \ \ \ \ \ \ \ \  \ \ \ \ \ \ \ \ \ \ \ \ \ \nonumber\\
    +\left(\frac{3}{2} \frac{\dot{\varphi}}{H}+6\xi\varphi - \frac{V'}{2H^2}\right)H\delta\varphi,\ \ \ \ \
\end{eqnarray}
with suitable boundary conditions. Furthermore, we obtain the
exact background dynamical equation
\begin{equation}\label{bg1}
    3H^2 (1-\xi\varphi^2)= -\frac{1}{2}\dot{\varphi}^2 + V,
\end{equation}
which coincides with expression (\ref{Fried mod1}) in the
slow-climb limit.

Now, in order to find the quadratic action for $\delta\varphi$, we
need to insert relations (\ref{sol 11}) and (\ref{sol 12}) in the
action (\ref{action2}) and expand it up to second order. However,
as we can see these expressions for $N$ and $N^i$  are subleading
in the slow-climb limit ($\dot{\varphi}^2\ll H^2$) and large-field
inflation ($\varphi^2\gg 1$) , comparing to $\delta\varphi$ (on
the other hand, if the momentum of the inflaton was comparable
with its energy density, namely $|\dot\varphi|\sim H$, the quantum
fluctuation of the background would become significant and could
cause instabilities on the background). Therefore, it is adequate
to consider just the action (\ref{action}) for $\delta\phi$ in the
de Sitter background, resulting in the second-order action
\cite{Feng:2009kb}
\begin{equation}
    S_2 = \frac{1}{2}\int d^4x a^3\bigg[ -\delta\dot{\varphi}^2 + (\nabla{\delta\varphi})^2 - V''\delta\varphi^2 -
    12\xi H^2\delta\varphi^2\bigg] .
\end{equation}
Moreover, introducing  the Fourier transform of $\delta\varphi$
through $\delta\varphi_k$, the perturbation equation writes
\begin{equation}
    \delta\ddot\varphi_k + 3H\delta\dot\varphi_k+\frac{k^2}{a^2}\delta\varphi_k  =0 \,,
\end{equation}
where we have used $\eta \ll 1$ and $\Delta\ll1$. Therefore, as we
observe, the quantum fluctuations in a Hubble time have the same
value as in the canonical case \cite{Guth:2007ng,Feng:2009kb}
\begin{equation}
\label{qunthubblr}
    \delta_q\varphi \approx \frac{H}{2\pi}.
\end{equation}

Expression (\ref{qunthubblr}) provides the quantum fluctuations of
the inflaton in one Hubble time. On the other hand, it is known
that usually the classical motion of the inflaton during one
Hubble time is given by \cite{Lindebook,LidseyRMP}
\begin{equation}
    |\delta_c\varphi| \approx |\dot{\varphi}H^{-1}| \sim \frac{|V'|}{3H^2}\bigg(1+\Delta\varphi^2\bigg).
\end{equation}
Thus, we deduce that if the quantum fluctuations are larger than
the classical ones, namely $\delta_q\varphi > |\delta_c\varphi|$,
then inflation will be eternal. Therefore, the necessary
conditions for exiting phantom inflation is to use the suitably
defined and bounded-from-above potentials of the phantom-inflation
literature
\cite{Piao:2004tq,Lidsey:2004xd,GonzalezDiaz:2004df,Nojiri:2005pu,Capozziello:2005tf,Wu:2006wu,Piao:2007ne,Elizalde:2008yf},
plus the condition $\delta_q\varphi < |\delta_c\varphi|$.
 Thus, since slow-climb always requires $\Delta\varphi^2\ll1$ and
 the validity of (\ref{Fried mod1}), the condition that prevent
 eternality reads
\begin{equation}\label{conditon00}
     \left|\frac{dV(\varphi)}{d\varphi}\right|\gtrsim
     \left|V(\varphi)\right|^{3/2}\left(1+\frac{3\xi}{2}\varphi^2\right).
\end{equation}
This condition restricts the potential-forms that can give rise to
a finite inflation, or inversely, for a given potential it
determines the bounds inside which the field can move, in order to
avoid eternality. Finally, in the limit $\xi\rightarrow0$ the
above relation provides the corresponding condition for the
minimal phantom inflation.

\section{Langevin analysis for the nonminimal slow-climb phantom scenario}
\label{Langanal}

In the previous section we extracted the general condition that prevents eternality in phantom inflation, estimating
separately the effects of the classical motion and of the quantum fluctuations. In this section we will try to verify
the aforementioned results, solving explicitly the cosmological equations, formulating the classical motion plus the
stochastic effect of the quantum fluctuations through a Langevin analysis \cite{Chen:2006hs}. In order to be able to
provide analytical results we will use the toy example of the exponential potential $V(\varphi)=V_0
e^{\lambda\varphi}$, with $\lambda>0$, which satisfies the basic requirements for phantom inflation.

The overall evolution of the phantom field, including quantum fluctuations, is modeled through a random walk, and
therefore it can be described by the following Langevin equation \cite{Chen:2006hs},
\begin{equation}\label{Langevin}
     3H \dot\varphi - V'(\varphi) - 12\xi H^2\varphi = \frac{3}{2\pi}H^{5/2}n(t)\,,
\end{equation}
where $n(t)$ is a Gaussian white noise normalized as
\begin{equation}\label{normal}
    \langle n(t)\rangle =0 \,,\qquad \langle n(t)n(t')\rangle = \delta(t - t')\,.
\end{equation}
As can be seen, $n(t)$ has dimensions of mass to the power of one half. Using the exponential potential, and taking the
approximation $3H^2\approx V_0$ during inflation, which means we do not consider the backreaction from the space-time
to the classical evolution of the inflaton and focus on the quantum fluctuation of the inflaton itself which is modeled
by the stochastic process, then we get
\begin{equation}\label{Langevin 2}
     \dot\varphi - \bigg(\lambda e^{\lambda\varphi}+4\xi\varphi\bigg)\sqrt{\frac{ V_0}{3}} =
     q
     n(t),
\end{equation}
where we have defined $q\equiv \frac{H^{3/2}}{2\pi}$.

In Eq.~(\ref{Langevin 2}), if the term on the right-hand side
is absent we recover the usual slow-climb equation of motion and
the inflaton will follow a classical trajectory $\varphi_{c}(t)$.
Therefore, we expand the field $\varphi(t)$ around its classical
value $\varphi_c(t)$ up to order $\mathcal{O}(q^2)$, namely
\begin{equation}\label{expand}
    \varphi(t)= \varphi_c(t) + q\varphi_1(t)+q^2\varphi_2(t)+\mathcal{O}(q^3) \,.
\end{equation}
Substituting this expansion into (\ref{Langevin 2}) and setting
the coefficients of the $q$-powers to zero, we acquire the
equations
\begin{eqnarray}\label{lang equs}
  \dot \varphi_c &=& \sqrt{V_0/3}\big(\lambda e^{\lambda\varphi_c} +4\xi\varphi_c\big) \\
  \dot \varphi_1 &=& \sqrt{V_0/3}\big(\lambda^2 e^{\lambda\varphi_c} +4\xi\big)\varphi_1 + n(t) \\
  \dot \varphi_2 &=& \sqrt{V_0/3}\bigg[\frac{1}{2}\lambda^3 e^{\lambda\varphi_c} \varphi_1^2+\big(\lambda^2 e^{\lambda\varphi_c} +4\xi\big)\varphi_2
  \bigg].
\end{eqnarray}
These three equations can be solved analytically in Case I and
Case II of (\ref{eomp21}),(\ref{eomp22}), namely for $
|f\varphi|\ll |V'|$ and $|f\varphi|\gg |V'|$ respectively. The
explicit solutions are presented in the Appendix.

For case I, the condition for the Hubble parameter not to be
changed significantly by the quantum noise (see Appendix
\ref{Appendix1}) reads
\begin{equation}
\label{cond bound0}
    \varphi_0\lesssim \lambda^{-1}\ln V_0^{-1} \,,
\end{equation}
while for Case II the corresponding condition (see Appendix
\ref{Appendix2}) reads
\begin{equation}\label{bound case20}
    \varphi_0 \lesssim \lambda^{-1}\ln V_0^{-1} + \lambda^{-1}\ln (\sqrt{\xi}/\lambda).
\end{equation}
In other words, if these conditions are satisfied, that is if the
inflaton remains smaller than these critical values, then
inflation will not be eternal.

Let us now compare these expressions with the condition
(\ref{conditon00}) derived in the previous section. Applying
(\ref{conditon00}) in the case of the exponential potential of the
present section, and keeping up to zeroth order in terms of $\xi$
(since otherwise we obtain transcendental equations), we acquire
\begin{equation}
 \varphi_0\lesssim \lambda^{-1}\ln V_0^{-1} + 2\lambda^{-1} \ln
 \lambda.
 \end{equation}
Clearly, this expression is consistent with both (\ref{cond
bound0}) and (\ref{bound case20}), and the slight differences
arise from the performed assumptions that were necessary in order
to solve the Langevin equation. Additionally, going to first order
in $\xi$ in (\ref{conditon00}), one can numerically show the
agreement too. Therefore, we conclude that the results of the
previous sections are indeed reliable.

\section{Conclusions}
\label{conclusions}

In this work we investigated the necessary conditions that prevent
phantom inflation to be eternal, going beyond the basic conditions
of slow-climb behavior. In particular, even using potentials
bounded from above and with suitable slopes, which give rise to
slow climbing, quantum fluctuations could still lead inflation to
be eternal. Thus, after presenting the slow-climb conditions, we
performed an analysis of the fluctuations, extracting the overall
conditions that are necessary for preventing eternality. Finally,
in order to be general, we moreover allowed for a nonminimal
coupling of the phantom field with gravity.

Our main result is expression (\ref{conditon00}), which is the
condition restricting the potential-forms that can give rise to a
finite inflation, or inversely the condition determining the
bounds inside which the field can move in a given, slow-climb
potential, in order to avoid eternality. Note that in our analysis
we did not need any additional mechanism in order to exit eternal
phantom inflation, such as the use of an extra scalar
\cite{Piao:2004tq}, the imposition of strong backreaction
\cite{Wu:2006wu}, the consideration of multiuniverses
\cite{GonzalezDiaz:2004df},  or the use of specially-designed
braneworld models with  brane/flux annihilation
\cite{Piao:2007ne}.

Furthermore, in order to verify the obtained results, we solved
explicitly the cosmological system in a simple example of an
exponential potential, formulating the classical motion plus the
stochastic effect of the quantum fluctuations through Langevin
equations. Requiring finite parameters in the inflation we resulted to similar
conditions with those obtained by the above fluctuation-analysis
procedure.

Let us make a comment here, on the limits of applicability of our
analysis. First of all, as we have mentioned, the phantom field
must be smaller than the Planck scale, thus its backreaction will
be small and not capable of bringing inflation to eternality (in
Langevin-equation terms, this means that the expansion around the
classical trajectory (\ref{expand}) is valid). However, in an
inflating universe, even if the examined region satisfies these
conditions, its neighboring regions can have very high densities,
and thus one could ask whether this behavior could bring about
strong quantum effects in the examined region too. Therefore, we
have to make an additional assumption, namely that the initially
low-density, slow-roll-inflating region has been already causally
disconnected from its possible high-density neighboring regions,
and the possible interactions lie outside the horizon. In such a
case, the inflation of the observable universe will not be led to
eternality.

Phantom fields could have interesting implications either in
inflation or in describing the late-time acceleration of the
Universe. Although their quantum behavior could be problematic at
first, one can consider the phantoms to arise through an effective
description of a nonphantom, fundamental, higher-dimensional,
underlying theory, consistently with the basic requirements of
quantum field theory. Therefore, the examination of their
cosmological implications is valuable and can improve our
understanding of nature. In these lines, the fact that phantom
inflation can be noneternal makes the scenario at hand a
candidate for the description of the early universe.

\begin{acknowledgments}
This work is supported by National Education Foundation of China grant No. 2009312711004 and Shanghai Natural Science
Foundation, China grant No. 10ZR1422000.
\end{acknowledgments}

\appendix

\section{Solution of the Langevin equations}\label{generic potential}

Since we are dealing with stochastic variables, we perform the
average of any physical quantity by defining the statistical
measure. In particular, we use the Fokker-Planck approach and
define the measure to be the physical volume of the Hubble patch,
and thus the average is defined as
\begin{equation}\label{average}
    \langle H(t)\rangle_p = \frac{\langle H(t)e^{3N(t)}\rangle}{\langle e^{3N(t)} \rangle} \,,\quad N(t) = \int_0^t H(t')dt' \,.
\end{equation}
Since the Hubble patch that is eternally inflating will have an exponentially larger physical volume, taking the
largest weight in the average at late times, the physical volume can be a good measure to characterize eternal
inflation. Therefore, the average $\langle H(t)\rangle_p $ could be significantly changed by quantum fluctuations if
eternal inflation is realized. Furthermore, we shall use the functional technique developed in \cite{Chen:2006hs} and
define a generating functional
\begin{equation}\label{generating fun}
    W_t[\mu] = \ln\langle e^{M_t[\mu]} \rangle \,, \quad M_t[\mu] = \int_0^t \mu(t')H(t') dt' \,.
\end{equation}
Thus, $\langle H(t)\rangle_p $ can be evaluated by functionally
differentiating $W_t[\mu]$ with respect to $\mu$ and setting
$\mu=3$, resulting to the following equations up to
$\mathcal{O}(q^2)$:
\begin{eqnarray}
  \langle H(t)\rangle_p  &=&  \frac{\delta W_t[\mu]}{\delta \mu(t)}
  \bigg|_{\mu(t)=3}
  \nonumber\\
  & =& \langle H(t)\rangle + 3\int_0^t \langle\langle H(t)H(t')\rangle\rangle dt' ,\ \ \ \ \ \\
  \langle\langle H(t)H(t')\rangle\rangle &=& \langle H(t)H(t')\rangle -\langle H(t)\rangle_p\langle H(t')\rangle_p \,.
\end{eqnarray}
After these definitions we can proceed to the solution of the
Langevin equations.

\subsection{Case I:  $ |f\varphi|\ll |V'|$}
\label{Appendix1}

In this case, the phantom field can be regarded as minimally
coupled to gravity and the solution to (\ref{lang equs}) writes:
\begin{eqnarray}
\label{phi1}
  \varphi(t) &=& \varphi_{c}(t) + q\, e^{\lambda\varphi_c(t)}\Xi(t) + q^2\, e^{\lambda\varphi_c(t)}\Pi(t')
   \end{eqnarray}
   with
   \begin{eqnarray}
  \varphi_c(t) &=& -\lambda^{-1} \ln\bigg[ e^{-\lambda\varphi_0} -\lambda^2t\sqrt{V_0/3}\bigg]\,,
\end{eqnarray}
where the subscript $0$ denotes the initial value of the field (at
$t=0$). In (\ref{phi1}) we have defined the quantities
\begin{eqnarray}
  \Xi(t) &=& \int_0^t n(t')e^{-\lambda\varphi_c(t')} dt' 
   \nonumber\\
  &=& \int_0^t n(t')\bigg[ e^{-\lambda\varphi_0} -\lambda^2t'\sqrt{V_0/3}\bigg]
    dt'
    \end{eqnarray}
    and
    \begin{eqnarray}
  \Pi(t) &=& \frac{\lambda^3}{2} \sqrt{\frac{V_0}{3}}\int_0^t e^{2\lambda\varphi_c(t')}\Xi^2(t')
  dt',
\end{eqnarray}
where $\Xi(t)$ is a new stochastic variable normalized as
\begin{eqnarray}
 && \langle \Xi(t) \rangle = 0 , \label{nor1}\\
 && \langle \Xi(t)\Xi(t')\rangle 
 \nonumber\\
  &&= \frac{ e^{-3\lambda\varphi_0}  }{\lambda^2\sqrt{3V_0}}
  \bigg\{1-
  \left[1-\lambda^2
  e^{\lambda\varphi_0}\sqrt{V_0/3}\text{min}(t,t')\right]^3\bigg\}. \ \ \ \, \ \ \ \ \label{nor2}
\end{eqnarray}
The Hubble parameter reads
\begin{eqnarray}\label{hubble lang}
    H(t) = H_c(t) + q\, \sqrt{\frac{V_0}{3}}\frac{\lambda}{2} e^{3\lambda\varphi_c(t)/2}
    \Xi(t)\ \ \ \ \ \ \ \ \ \ \ \ \ \ \nonumber\\
    + q^2\, \sqrt{\frac{V_0}{3}}\frac{\lambda}{8} e^{3\lambda\varphi_c(t)/2}\left[\lambda e^{\lambda\varphi_c(t)}\Xi^2(t) + 4\Pi(t) \right], \
\end{eqnarray}
where $H_c(t) =
\sqrt{V(\varphi_c(t))/3}=e^{\lambda\varphi_c/2}\sqrt{V_0/3}$.
Using (\ref{nor1}), (\ref{nor2}) and (\ref{hubble lang}), we can
further obtain
\begin{equation}\label{hubble lang ave}
    \langle H(t)\rangle = \langle H_c(t)\rangle + q^2\, \frac{e^{-2\lambda\varphi_0}}{8}e^{3\lambda\varphi_c(t)/2}\bigg(e^{\lambda[\varphi_c(t)-\varphi_0]}
    -1\bigg)
\end{equation}
and
\begin{eqnarray}
   3\int_0^t \langle\langle H(t)H(t')\rangle\rangle = \frac{q^2e^{-\lambda\varphi_c(t)}}{10\lambda^2}
    \left\{5e^{3\lambda[\varphi_c(t)-\varphi_0]}\right.
    \nonumber\\
    \left.+1-6e^{5\lambda[\varphi_c(t)-\varphi_0]/2}\right\}.
\label{hubble lang ave2}
\end{eqnarray}

Now, in the limit $t\ll t_0
\equiv\lambda^{-2}e^{-\lambda\varphi_0}\sqrt{3/V_0}$ we can
acquire the leading order behavior of $\langle H(t)\rangle_p$ in
terms of $t$ as
\begin{eqnarray}\label{hubble lang ave3}
  \langle H(t)\rangle_p = \langle H(t=0)\rangle_p\ \ \ \ \ \ \ \ \ \  \ \ \ \ \  \ \  \ \ \ \ \  \ \ \ \ \ \ \  \ \ \ \ \  \ \ \nonumber\\
 + \frac{e^{\lambda\varphi_0/2}}{2}\sqrt{\frac{V_0}{3}}\left(\frac{t}{t_0}\right) +
   \frac{q^2e^{-\lambda\varphi_0/2}}{8}\left(\frac{t}{t_0}\right),\
   \
\end{eqnarray}
where the second term arises from expanding the classical motion
$H_c(t)$, while the last term comes from the quantum correction
(\ref{hubble lang ave}). Note that the contribution from
 (\ref{hubble lang ave2}) is of the order of $(t/t_0)^2$.
Requiring the Hubble parameter not to be changed significantly by
the quantum noise, we need to impose
\begin{equation}\label{cond}
   \frac{e^{\lambda\varphi_0/2}}{2}\sqrt{\frac{V_0}{3}} \lesssim \frac{q^2e^{-\lambda\varphi_0/2}}{8},
\end{equation}
which provides the bound when $(t/t_0)\ll1$ as:
\begin{equation}\label{cond bound}
    \varphi_0\lesssim\lambda^{-1}\ln V_0^{-1}.
\end{equation}

\subsection{Case II:  $ |f\varphi|\gg |V'|$}
\label{Appendix2}

In this case, the solution to (\ref{lang equs}) reads
\begin{eqnarray}
  \varphi(t) &=& \varphi_c(t) + q\,\varphi_c(t)\Xi(t) + q^2\,\varphi_c(t) \frac{\varphi_{20}}{\varphi_0}
    \end{eqnarray}
   with
   \begin{eqnarray}
  \varphi_c(t) &=&\varphi_0 \exp\bigg(4\xi t \sqrt{V_0/3}\bigg),
\end{eqnarray}
where $\varphi_{20} \equiv \varphi_2(t=0)$ and similarly to the
previous subsection we have defined
\begin{eqnarray}
  \Xi(t) &=& \int_0^t n(t')\varphi_c^{-1}(t') dt' 
  \nonumber\\
  &=& \varphi_0^{-1}\int_0^t n(t') \exp\bigg(-4\xi t \sqrt{V_0/3}\bigg)
    dt' ,
\end{eqnarray}
normalized as
\begin{eqnarray}
 && \langle \Xi(t) \rangle = 0 , \label{nor11}\\
 && \langle \Xi(t)\Xi(t')\rangle
 \nonumber\\
  &&= \frac{\varphi_0^{-2}}{8\xi\sqrt{V_0/3}} \left\{1-\exp\left[-8\xi \sqrt{V_0/3} \text{min}(t,t')\right] \right\}.\ \ \ \ \ \  \label{nor22}
\end{eqnarray}
The Hubble parameter is
\begin{eqnarray}
    H(t) = H_c(t) + q\, \sqrt{\frac{V_0}{3}}\frac{\lambda}{2} e^{\lambda\varphi_c(t)/2} \varphi_c(t)
    \Xi(t)\ \ \ \ \ \ \ \ \ \ \ \ \ \ \nonumber\\
    + q^2\, \sqrt{\frac{V_0}{3}}\frac{\lambda}{8} e^{3\lambda\varphi_c(t)/2}
    \left[\lambda \varphi_c^2(t)\Xi^2(t) + 4\varphi_c(t) \frac{\varphi_{20}}{\varphi_0}
    \right],\ \ \ \,
    \label{hubble lang case 2}
\end{eqnarray}
where $H_c(t) = e^{\lambda\varphi_c/2}\sqrt{V_0/3}$. Moreover we
obtain
\begin{eqnarray}
  \langle H(t)\rangle = \langle H_c(t)\rangle + q^2\, \sqrt{\frac{V_0}{3}}\frac{\lambda}{8} e^{3\lambda\varphi_c(t)/2}
\left\{  4\varphi_c(t)
\frac{\varphi_{20}}{\varphi_0}\right.\nonumber\\
\left. + \frac{\lambda}{8\xi\sqrt{V_0/3}}
\left[\left(\frac{\varphi_c(t)}{\varphi_0}\right)^2-1\right]
    \right\}\ \ \ \ \
    \label{hubble lang ave case2}
\end{eqnarray}
and
\begin{eqnarray}
 && 3\int_0^t \langle\langle H(t)H(t')\rangle\rangle = \ \ \ \  \ \ \ \
  \ \ \  \ \ \ \  \ \ \ \  \  \ \  \ \ \  \ \ \ \  \ \ \ \  \nonumber \\
&&\  \ \ \ \  \    =
\frac{3q^2\lambda^2e^{\frac{\lambda\varphi_c}{2}}}{128\xi^2}
\left(\frac{\varphi_c}{\varphi_0}\right)
    \left\{\frac{2}{\lambda\varphi_0}\left(e^{\frac{\lambda\varphi_c}{2}}-e^{\frac{\lambda\varphi_0}{2}}\right)\right.\nonumber\\
  && \  \ \ \ \  \        +\left(\frac{\varphi_0}{\varphi_c}e^{\frac{\lambda\varphi_c}{2}}-e^{\frac{\lambda\varphi_0}{2}}\right)\nonumber\\
    &&  \  \ \ \ \  \     \left.
         +\frac{\lambda\varphi_0}{2}\left[\text{Ei}\left(-\frac{\lambda\varphi_c}{2}\right)-\text{Ei}\left(-\frac{\lambda\varphi_0}{2}\right)\right]
    \right\},
    \label{hubble lang ave2 case2}
\end{eqnarray}
where Ei is the exponential integral function.

In the limit $t\ll t_0 \equiv\sqrt{3/V_0}/(4\xi)$ we can obtain
the leading order behavior of $\langle H(t)\rangle_p$ in terms of
$t$ as
\begin{eqnarray}
   \langle H(t)\rangle_p = \langle H(t=0)\rangle_p + \frac{\lambda\varphi_0e^{\lambda\varphi_0/2}}{2}\sqrt{\frac{V_0}{3}}\left(\frac{t}{t_0}\right)
   \ \ \ \  \  \ \ \ \  \  \ \ \  \nonumber\\
   +q^2 \sqrt{\frac{V_0}{3}}\frac{\lambda}{8} e^{3\lambda\varphi_0/2}
    \bigg[\frac{\lambda}{4\xi\sqrt{V_0/3}}+ 4\varphi_{20}+ 6\lambda\varphi_0\varphi_{20}\bigg]\left(\frac{t}{t_0}\right), \ \
     \nonumber
\end{eqnarray}
where the second term arises from expanding the classical motion
$H_c(t)$ and the last term comes from the quantum correction
(\ref{hubble lang ave case2}). Note that the contribution from
(\ref{hubble lang ave2 case2}) is of the order of $(t/t_0)^2$.
Requiring the Hubble parameter not to be changed significantly by
the quantum noise, we impose
\begin{equation}\label{cond case2}
   4\varphi_0
\lesssim \frac{q^2\,  \lambda e^{\lambda\varphi_0}
}{4\xi\sqrt{V_0/3}} \,,
\end{equation}
where we have used that $\xi\ll1$ and
$\varphi_0\gg\Delta^{-1/2}\sim\xi^{-1/2}$. Thus, we conclude that
at $(t/t_0)\ll1$:
\begin{equation}\label{bound case2}
    \varphi_0 \lesssim \lambda^{-1}\ln V_0^{-1} + \lambda^{-1}\ln
    (\sqrt{\xi}/\lambda).
\end{equation}

\end{document}